\pdfoutput=1
\documentclass[runningheads]{llncs}

\usepackage{amssymb}
\usepackage{graphicx}

\usepackage{url}

\usepackage{cite,epsfig,verbatim,rotating,subfigure}
\usepackage{times}
\usepackage{url}
\usepackage{algorithm2e}
\usepackage{amsfonts}

\begin{document}

\mainmatter  

\title{Multiple Sequence Alignment System for Pyrosequencing Reads}

\titlerunning{Multiple Sequence Alignment System for Pyrosequencing Reads}

%
%
%

\author{Fahad Saeed$^{1}$ \and Ashfaq Khokhar$^{1}$ \and Osvaldo Zagordi$^{2}$ \and Niko Beerenwinkel$^{2}$}

\authorrunning{Saeed Khokhar Zagordi and Beerenwinkel}


\institute{$^{1}$Department of Electrical and Computer Engineering,\\
University of Illinois at Chicago, IL USA\\
and\\
$^{2}$Department of Biosystems Science and Engineering \\
ETH Zurich, Basel, Switzerland\\
}

%
%

\maketitle

\begin{abstract}
Pyrosequencing is among the emerging sequencing techniques,
capable of generating upto 100,000 overlapping reads in a single
run. This technique is much faster and cheaper than the existing
state of the art sequencing technique such as Sanger. However, the
reads generated by pyrosequencing are short in size and contain
numerous errors. Furthermore, each read has a specific position in
the reference genome. In order to use these reads for any
subsequent analysis, the reads must be  aligned . Existing
multiple sequence alignment methods cannot be used as they do not
take into account the specific positions of the sequences with
respect to the genome, and are highly inefficient for large number
of sequences. Therefore, the common practice has been to use
either simple pairwise alignment despite its poor accuracy for
error prone pyroreads, or use computationally expensive techniques
based on sequential gap propagation. In this paper, we develop a
computationally efficient method based on domain decomposition,
referred to as {\it pyro-align}, to align such large number of
reads. The proposed alignment algorithm accurately aligns the
erroneous reads in a short period of time, which is orders of
magnitude faster than any existing method. The accuracy of the
alignment is confirmed from the consensus obtained from the
multiple alignments.


\end{abstract}

\section{Introduction}

\label{sec:intro}

Pyrosequencing is among the emerging sequencing techniques
developed for determining the sequences of DNA bases from a
genome. It is capable of generating up to 100,000 overlapping
reads in a single run. However, multitude of factors, such as
relatively short read lengths (i.e., as of 2008 an average of
$100-250$ nt compared to $800-1000$ nt for Sanger sequencing),
lack of a paired end protocol, and limited accuracy of individual
reads for repetitive DNA, particularly in the case of monopolymer
repeats, present many computational challenges~\cite{dna} to make
pyrosequencing useful for biology and bioinformatics applications.

For over more than a decade, Sanger sequencing has been the
cornerstone of genome sequencing including that of microbial
genomes. Improvements in DNA sequencing techniques and the
advances in data storage and analysis, as well as developments in
bioinformatics have reduced the cost to a mere $8000\$-10000\$$
per megabase of high quality genome draft sequence. However, the
need of more efficient and cost effective approaches has led to
development of new sequencing technologies such as the 454 GS20
sequencing platform. It is a non-cloning pyrosequencing based
platform that is several orders of magnitude faster than the
Sanger machines. However, the new technology despite its enormous
advantage in terms of time and money will not be able to replace
the current Sanger technology, unless the reads generated are
properly aligned with respect to the reference genome.

The key issues associated with the use of pyrosequencing technique
are as under:

\emph{Read Length:} The read length is expected to be of the order
of $100-250bp$ on average. This is much shorter than the other
state of the art Sanger machines which give out consistent read
lengths of the order of $> 800-900bp$.

\emph{Orientation:} This is generally the case for most of the
sequencing technologies. Each DNA helix will be broken into the
original and its Watson-Crick complement. These would be further
broken up into pieces, and there is generally no way to reveal
which of the two is it. The problem is more severe and usually
encountered for genome reconstruction.

\emph{Errors:} Each individual DNA sequence or read is likely to
have errors in the form of insertions and deletions. It may also
have mutations and the pyrosequencer may itself make errors. These
errors correspond to homopolymer effects, including extension
(insertions), incomplete extensions (deletions), and carry forward
errors (insertions and substitutions). Insertions are considered
the most common type of error ($36\%$ of errors) followed by
deletions ($27\%$), ambiguous bases, Ns ($21\%$), and
substitutions ($16\%$) \cite{massive}.

\begin{figure}[htb]
\vspace{-12mm}
\begin{center}
\includegraphics[scale=0.4,angle=90]{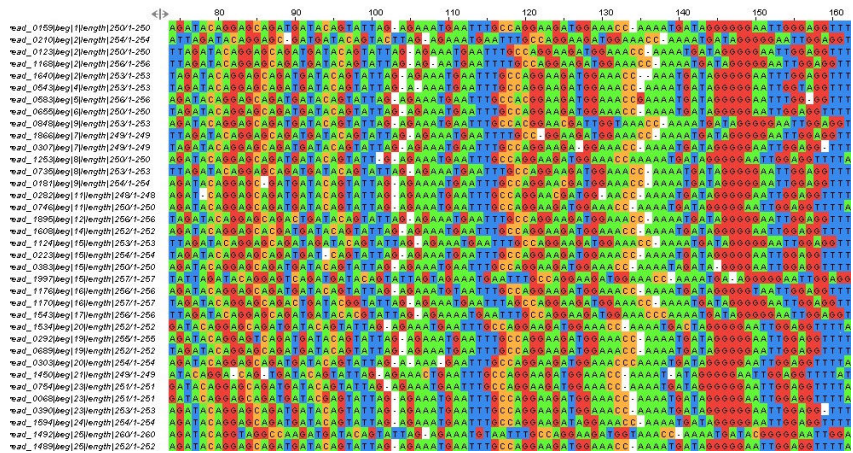}\vspace{-8mm}
\caption{\small \label{fig-pairwise} Pairwise alignment of the
reads with the reference genome is shown} \vspace{-8mm}
\end{center}
\end{figure}

For most practical purposes, pyroreads without any post processing
are of limited use. One of the most widely required tasks as a pre
processing step for many applications, including haplotype
reconstruction \cite{niko}\cite{niko2}, analysis of microbial
community analysis \cite{zong}, analysis of genes for
diseases\cite{xl}, is the alignment of these reads with the wild
type. For important applications such as viral population
estimation or haplotype reconstruction of various viruses e.g.,
HIV in a population, scientists usually have the information about
the wild type genome of the virus. While for other sequencing
technologies, such as Sanger, simple pair-wise alignment with the
wild type may produce reasonable multiple alignment, in the case
of pyrosequencing, the variation in the haplotype population
compounded with the errors introduced in the reads does not allow
feasible multiple alignment by simple pair-wise alignment.
Fig.~\ref{fig-pairwise} depicts simple pair-wise alignment of
pyrosequence reads with a reference genome. We assert that
accurate and workable multiple alignment is often necessary for a
variety of applications and statistical packages to work with
these pyroreads, as demonstrated in
\cite{niko}\cite{niko2}\cite{zong}\cite{xl}.

In theory, alignment of multiple sequences can be achieved using
pair-wise alignment, each pair getting alignment score. But for
optimal alignment the sum of all the pair-wise alignment scores
need to be maximized, which is an NP complete
problem~\cite{NP-hard}. Towards this end, dynamic programming
based solutions of $O(L^N)$ complexity have been pursued, where
$N$ is the number of sequences and $L$ is the average length of a
sequence. Such accurate optimizations are not practical for large
number of sequences -as is the case in pyrosequencing- , thus
making heuristic algorithms as the only feasible option. The
literature on these heuristics is vast and includes widely used
works, such as Notredame et. al.~\cite{tcoffee}, Edgar
~\cite{Muscle}, Thompson et. al.~\cite{clustalw}, Do et. al.
~\cite{probcons}, and Morgenstern et.al.~\cite{dialign}. These
heuristics are complex combination of ad-hoc procedures with some
flavor of dynamic programming. Despite the usefulness of these
widely used heuristics, they scale very poorly with increasing
number of sequences.

For multiple alignment of pyroreads, 'out of the box' use of these
heuristics is not feasible because of two main reasons: 1) the
pyrosequencing reads can be very large in number (up to $100,000$
usable reads in a single run (with a Roche GS20 platform), and 2)
the heuristics do not take into account the positions of the reads
with respect to the reference genome. Additional factors such as
short lengths and errors, and the fact that these reads have
preceding or trailing 'gaps' pose further alignment challenges. In
~\cite{niko}, an alignment technique based on sequential gap
propagation has been used. This technique is computationally
expensive and its alignment quality decreases with the increase in
the mutation value.

In this paper, we present a computationally efficient algorithm
\emph{pyro-align}, specifically designed for multiple alignment of
DNA reads obtained from pyrosequencing. The proposed algorithm is
based on a novel domain decomposition concept, therefore it is
capable of aligning very large number of pyrosequences. It takes
into account the position of the reads with respect to the
reference genome, and assigns weight to the leading and trailing
gaps for the reads.

The objective of our work is to develop a multiple alignment
system for small error prone reads, such that the errors in the
alignment are 'highlighted' and the system is able to handle large
number of reads, as may be expected from pyrosequencing reads.

We assume that the reads may be generated from one or many
genomes, with 'forward' orientation. We also assume that the
reference genome (or its wild type) from which the reads are
generated is available, as is generally the case for haplotype
reconstruction. In our experiments, we have used HIV-pol gene
virus as the reference genome (with length of 1970bp) and
simulator Readsim~\cite{readsim} to generate these reads. The
algorithm uses concepts from domain decomposition and parallel
multiple alignment techniques \cite{SaeedKhokhar, SaeedKhokhar2}.

For the sake of completeness, let's first formally define the
Multiple Sequences Alignment problem in its generic form, without
indulging with the issues such as scoring functions. Let $N$
sequences be presented as a set $S = \{S_1,S_2, S_3, \cdots,
S_N\}$ and let $S^{'} = \{ S^{'}_{1}, S^{'}_{2} , S^{'}_{3},
\cdots, S^{'}_{N} \}$ be the aligned sequence set, such that all
the sequences in $S^{'}$ are of equal length, have maximum
overlap, and the score of the global map is maximum according to
some scoring mechanism suitable for the application.

A perfect multiple alignment for pyroreads would be, that the
reads are aligned with each other such that the position of the
reads with respect to the reference genome is conserved; the reads
have maximum overlap and are of equal lengths after the alignment,
including leading and trailing gaps.



The intuitive idea behind the proposed {\it pyro-align} algorithm
is  to first place the reads in correct orientation with respect
to the reference genome and then use progressive alignment to
achieve the final alignment. For efficient progressive alignment,
the correctly placed reads are reordered according to the starting
position, and a computationally low complexity similarity metric
is extracted from this ordering position. The similarity metric is
then used to align pairs of aligned reads using a hierarchical
decomposition strategy.
 The proposed multiple alignment algorithm takes advantage of the
 pyroreads characteristics and brings in techniques from data
 structures and parallel computing to realize a  low complexity
 solution in terms of time and memory.

The proposed alignment algorithm, {\it pyro-align}, consists of
the following two main components:

\begin{enumerate}
\item Semi-Global alignment
 \item Hierarchical progressive alignment
\begin{enumerate}
\item Reordering of reads to generate guidance tree
 \item Pairwise and profile-profile alignment
\end{enumerate}
\end{enumerate}

Each component is designed considering the characteristics of
pyroreads and it is described in the following sections along with
its justification.

\subsection{Semi-Global Alignment}
The first step is to determine the position of each read with
respect to the reference genome. If this step is omitted, there
are number of alignments that would be correct, but would be
inaccurate if analyzed in the global context. A read that is not
constricted in terms of position, may give the same score (SP
score) for the multiple alignment but would be incorrect in
context of the reference. To accomplish the task of 'placing' the
reads in the correct context with respect to the reference genome
we employ semi-global alignment procedure.

The semi global alignment is also referred to as overlapping
alignment because the sequences are globally aligned ignoring the
start and end gaps. For semi-global alignment we use a modified
version of Needleman-Wunsch algorithm~\cite{needle}.


The modification in the basic version of Needleman-Wunsch is
required to handle the leading and trailing gaps of the reads when
aligning to the reference genome. If the leading and trailing gaps
are not ignored, considering the short length of the reads, the
alignment scores would be dominated by these gaps, hence giving an
inaccurate alignment with respect to the genome.

Let the two sequences to be aligned be $s$ and $t$, and $M(i,j)$
presents the score of the optimal alignment. Since, we do not wish
to penalize the starting gaps, we modify the dynamic programming
matrix by initializing the first row and first column to be zero.
The gaps at the end are also not to be penalized. Let $M(i,j)$
represent the optimal score of $s_{1},\cdots,s_{i}$ and
$t_{1},\cdots,t_{j}$. Then $M(m,j)$ is the score that represents
optimally aligning $s$ with $t_{1,\cdots,j}$. The optimal
alignment therefore, is now detected as the maximum value on the
last row or column. Therefore the best score is $M(i,j)=max_{k,l}
(M(k,n),M(m,l))$, and the alignment can be obtained by tracking
the path from $M(i,j)$ to $M(0,0)$. For additional details on
semi-global alignment we refer the reader to \cite{comp-book}.

Once each read has been semi-globally aligned with the reference
genome, we obtain reads with leading and trailing gaps, where the
first character after the gaps is the starting position of the
read with respect to the reference genome. The information for
these alignments are stored in hashtables that are further used
for processing in reordering the reads for alignment.

\section {Hierarchical Progressive Alignment}

Generally multiple sequence alignment (MSA) procedures are either
based on iterative methods or employ progressive techniques.
Although, progressive techniques relative to iterative techniques
are more efficient, they are not suitable when the sequences are
relatively diverse or the number of sequences is very large.
Considering the fact that the pyroreads are highly similar, we
develop a hierarchical progressive alignment procedure that is
also computationally efficient for large number of reads.

Progressive alignment techniques develop final MSA by combining
pair-wise alignments beginning with the most similar pair and
progressing to the most distantly related. All progressive
alignment methods require two stages: a first stage in which the
relationships between the sequences are represented as a tree,
called a guide tree, and a second step in which MSA is built by
adding the sequences sequentially to the growing MSA according to
the guide tree. In the following, we describe the low complexity
components of {\it pyro-align}.

\subsection{Reordering Reads}

The method followed by most of the progressive multiple alignment
algorithms is that a quick similarity measure is computed that is
based on k-mer counting~\cite{kmer} or some other heuristic
mechanism. These pair-wise similarity measures (distances) are
tabulated in a matrix form
 and a tree is
constructed from this distance matrix using UPGMA or neighboring
joining. The progressive alignment is thus built, following the
branching order of the tree, giving a multiple alignment. These
steps require $O(N^2)$ time each, where $N$ is the number of
reads. To reduce this complexity, we exploit the fact that the
reads are coming from the same reference or nearly the same
reference. This in turn implies that the reads starting from the
same or near same 'starting' point with respect to the reference
genome are likely to be similar to each other. Therefore, we
already have the ordering information or the 'guide tree' from the
first step of the algorithm. Our guide tree, or the order in which
sequences will be aligned in the progressive alignment is from the
starting position of the reads from the first stage. Of course the
decomposition of the reads (the subtree of the profiles that we
built) doesn't render the reads in the same order as in
traditional progressive alignment, but nevertheless the order is
more or less the same when the profiles of these reads are
aligned.

Let there be $N$ number of reads $R={R_1,R_2,\cdots,R_N}$
generated from pyrosequencing technique, from the reference genome
of length $L_g$. Also, let the length of each read denoted by
$L(R)_p$. After executing semi-global alignment using the
algorithm discussed in the previous section, let each read be
presented by $R_{pq}$, where the $p^{th}$ read has $q$ leading
gaps and $L_g-q-L(R)_p$ trailing gaps. Then the reordering
algorithm would reorder the reads such that after the reads are
reordered using the information from the leading gaps, the read
$R_{pq}$ comes in ordering 'before' $R_{p(q+1)}$, $\forall$ $p,q
\in L_g$.


To execute the reordering in an efficient manner, we employ
hashtables that speed up the search process. We create two
hashtables: $hashtable_1$ uses fasta sequence tag as the hash key
and stores the corresponding starting position of the read;
$hashtable_2$ stores the read names (fasta sequence tag) and the
dna sequence it is associated with. Using these tables, the reads
are reordered in the database in linear time.

\subsection{Pair-wise and Profile-Profile Alignments}
\begin{figure}[htb]
\vspace{-4mm}
\begin{center}
\includegraphics[scale=0.3,angle=90]{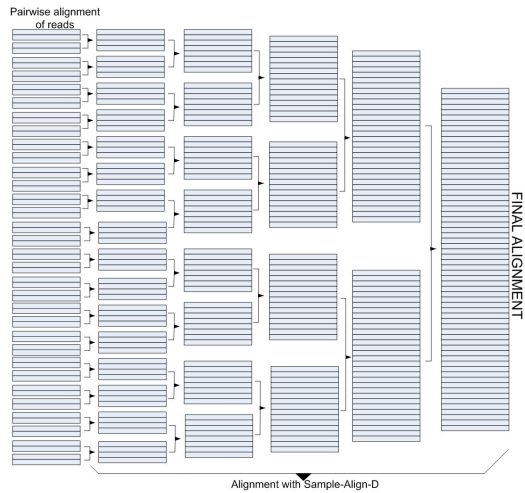}
\caption{\small \label{fig-samplealign} Hierchical profile-profile
alignments for pyro-align is shown  } \vspace{-8mm}
\end{center}
\end{figure}

The ordering of the reads determined in the preceding step is now
used to conduct the progressive alignment. Traditional progressive
alignment requires that the sequences most similar to each other
are aligned first. Thereafter, sequences are added one by one to
the multiple alignments determined according to some similarity
metric. This sequential addition of sequences for progressive
alignment is not suitable for large number of sequences. In order
to devise a low complexity system, we  design a hierarchical
progressive alignment procedure that is based on domain
decomposition~\cite{SaeedKhokhar}, as described below and depicted
in Figure~\ref{fig-samplealign}.

First of all, pair-wise local alignment using standard
Needle-Wunsch is executed on each overlapping pair of reads (the
ordering is still the same as discussed in the previous section).
After this stage, the reads are aligned in pairs such that we have
$N/2$ pairs of aligned reads. These $N/2$ pairs of reads are then
used for profile alignments as discussed below.

Profile-profile alignments are used to re-align two or more
existing alignments(in our case the pairs of aligned reads). It is
useful for two reasons; one being that the user may want to add
sequences gradually, and second being that the user may want to
keep one high quality profile fixed and keep on adding sequences
aligned to that fixed profile~\cite{clustalw}.

We take advantage of both of these properties in our domain
decomposition.

In this stage of the algorithm, the $N/2$ pairs of aligned reads
have to be combined to get a multiple alignment. We have shown in
\cite{SaeedKhokhar2} that the decomposition of the profiles gives
a fair amount of time advantages even on a single processor.
Therefore a hierarchical model similar to \cite{SaeedKhokhar} is
implemented (see Fig.~\ref{fig-samplealign}). The model requires
that instead of combining the profiles in a sequential manner (one
by one), a binary tree is built such that the profiles to be
aligned are the leafs of the tree.

\begin{figure}[htb]
\vspace{-25mm}
\begin{center}
\includegraphics[scale=0.4,angle=+90]{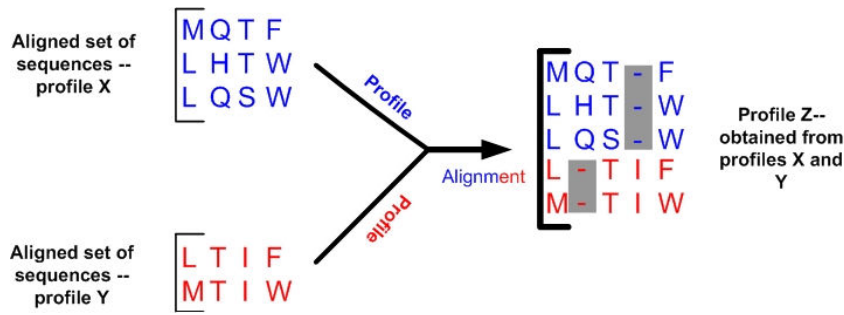}\vspace{-16mm}
\caption{\small \label{fig-profile} Two profiles(X and Y) are
aligned under the columns constrains, producing profile Z}
\vspace{-8mm}
\end{center}
\end{figure}

In order to apply pair-wise alignment functions to profiles, a
scoring function must be defined, similar to the substitution
methods defined for pair-wise alignments. One of the most commonly
used profile functions is the sequence-weighted sum of
substitution matrix scores for each pair of amino acid letters.
Let $i$ and $j$ be the amino acid, $p_i$ the background
probability of $i$, $p_{ij}$ the joint probability of $i$ and $j$
aligned to each other, $S_{ij}$ the substitution matrix being
used, $f^x_i$ the observed frequency of $i$ in column $x$ of the
first profile, $x_G$ the observed frequency of gaps in that
column. The same attributes are assumed for the profile $y$.
Profile sum of pairs (PSP) is the function used in Clustalw
\cite{clustalw}, Mafft \cite{mafft} and Muscle \cite{Muscle2} to
maximize Sum of Pairs(SP) score, which in turn maximizes the
alignment score such that the columns in the profiles are
preserved, as depicted in Fig.~\ref{fig-profile}.The PSP score can
be defined as in ~\cite{amino} and \cite{Muscle2}:

\begin{equation}
S_{ij}=\emph{log} (p_{ij}/p_i p_j)
\end{equation}

\begin{equation}
PSP^{xy}= \sum_{i}\sum_{j} f^{x}_{i} f^{y}_{j} \emph{log}
(p_{ij}/p_i p_j)
\end{equation}

\begin{figure}[htb]
\vspace{-20mm}
\begin{center}
\includegraphics[scale=0.4,angle=+90]{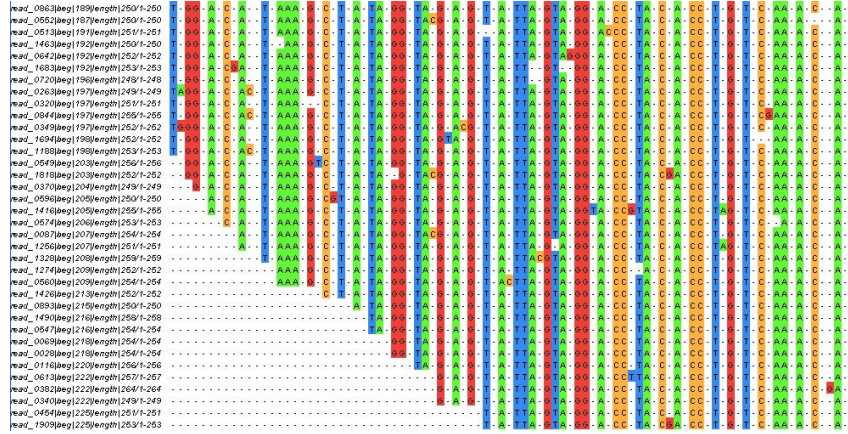}\vspace{-12mm}
\caption{\small \label{fig-aligned} The final Alignment of the
reads} \vspace{-8mm}
\end{center}
\end{figure}

For our purposes, we will take advantage of PSP functions based on
200 PAM matrix \cite{200PAM} and the 240 PAM VTML matrix
\cite{VTML}. Some multiple alignment methods implement different
scoring functions such as Log expectation (LE) functions, but for
our purposes PSP scoring suffices. Profile functions have evolved
to be quite complex and good discussion on these can be found at
\cite{Muscle2} and \cite{Muscle3}. We use the profile functions
from the clustalw system. The final alignment from the {\it
pyro-align} algorithm can be seen in Fig.~\ref{fig-aligned}.
Different steps of the proposed \emph{pyro-align} Algorithm are
outlined below.

\restylealgo{boxed}
\begin{algorithm}[H]
\SetLine \KwIn{Reads generated from pyrosequencing procedure and
Reference Genome} \KwOut{A Multiple Alignment of Reads is
returned}

//Calculate overlapping of each of the reads with respect to the reference Genome

\For{($i=1$;$i \leq N$;$i++$)}{ Overlapped-Reads $\leftarrow$
Semi-Global-Alignment($R_i$,Genome) \; }

Reordered-Reads $\leftarrow$ Reordering(Overlapped-Reads) \;

//Pairwise alignment using standard Needle-Wunsch is exectued, for pairs of ordered reads \;
Pair-wise-aligned $\leftarrow$ Needle-Wunsch(Reordered-Reads) \;

//Profile-profile alignment is obtained using Sample-align-D
strategy

Final-Alignment
$\leftarrow$Profile-Profile-alignment(Pair-wise-aligned) \;

\Return Final-Alignment \;

\caption{Steps of the Proposed Multiple Sequence Alignment
\emph{pyro-align} Algorithm}
\end{algorithm}


\section{Performance Analysis}\label{performance}

As discussed earlier in the paper, the exact solution for multiple
alignment is not feasible and heuristics are employed. Most of
these heuristics perform well in practice but there is generally
no theoretical justification possible for these heuristics
\cite{Gusfield}. For pyro-align it can be shown that the
semi-global alignment of the reads with the reference genome is
analogous to center star alignment. The center star alignment is
shown to give results within 2-approx of the optimal alignment
\cite{Gusfield} in worst case and same can be expected from the
semi-global alignment of reads with reference genome. The accuracy
of the later stages is confirmed by rigorous quality assessment
procedure described in the section below.

\subsection{Experimental Setup and Quality Assessment}
The performance evaluation of the algorithm has been  carried on a
single desktop system 2x QuadCore Intel 5355 2.66 GHz, 2x4 MB
Cache and 16GB of RAM. The operating system on the desktop is
RedHat Linux with kernel 2.6.18-92.1.13.el5. The software uses
libraries from Biojava \cite{biojava} and is built using java
version "1.6.0" Java(TM) SE Runtime Environment,IBM J9 VM.

\begin{figure}[htb]
\begin{center}
\includegraphics[scale=0.4,angle=90]{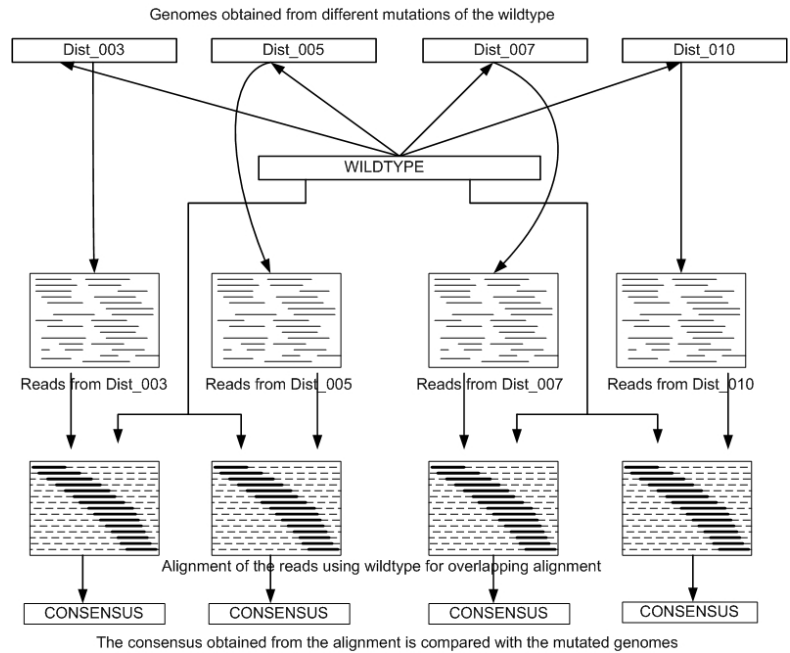}
\caption{\small \label{fig-quality} The experimental setup for the
quality assessment of the multiple alignment program}
\vspace{-8mm}
\end{center}
\end{figure}

To investigate the quality of the alignment produced by the
algorithm we used Readsim simulator \cite{readsim} to generate the
reads. The quality assessment of multiple alignment is generally
carried out using benchmarks such Prefab\cite{Muscle} or
BaliBase\cite{balibase}. However, these benchmarks are not
designed to access the quality of the aligned reads produced from
pyrosequencing, and there are no benchmarks available specifically
for these reads. Therefore, a system has to be developed to access
the quality of the aligned reads. The experimental setup for the
quality assessment of the alignment procedure is shown in the
Fig.~\ref{fig-quality} and is explained below.

Our quality assessment have two objectives: $(1)$to assess the
quality of the alignment produced by pyro-align with respect to
the original genome $(2)$ ensure that the system must be able to
handle reads from multiple haplotype for alignment.

To achieve these objectives, we setup the quality assessment
system as shown  Fig.~\ref{fig-quality}. We used a HIV pol gene
virus with length of 1970bp as the wildtype for the experiments.
The wildtype is then used to produce 4 sets of genomes, randomly
mutated at different rate; The four sets of genomes are Dist-003,
Dist-005, Dist-007 and Dist-010, with mutations of 3$\%$,
5$\%$,7$\%$ and 10$\%$, respectively. Now using the mutated
genomes, 2000 and 5000 reads from the Readsim were generated using
standard ReadSim parameters with forward orientation.

The generated reads from these mutated genomes were then aligned
with the wildtype.This procedure is adopted because generally
scientists only have a wildtype of the microbial genomes available
and therefore it depicts a more practical scenario.

\begin{figure}[htb]
 \vspace{-12mm}
\begin{center}
\includegraphics[scale=0.4,angle=0]{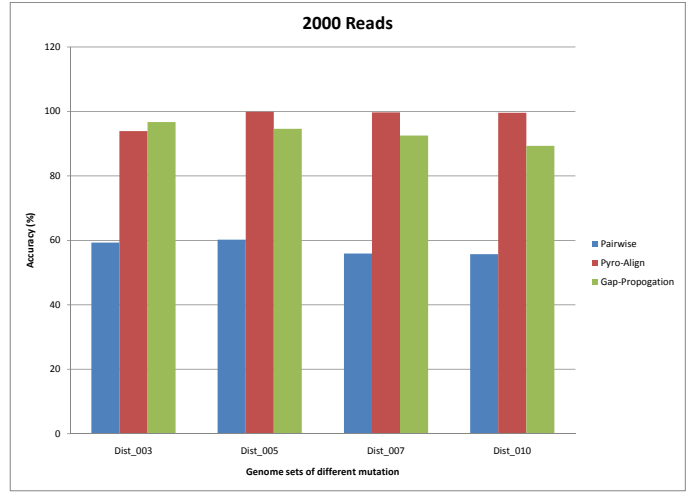} \vspace{-8mm}
\caption{\small \label{fig-2000} The quality of the alignment
using pairwise, pyro-Align and 'propagation' methods for 2000
reads}
 \vspace{-8mm}
\end{center}
\end{figure}

\begin{figure}[htb]
\vspace{-12mm}
\begin{center}
\includegraphics[scale=0.4,angle=0]{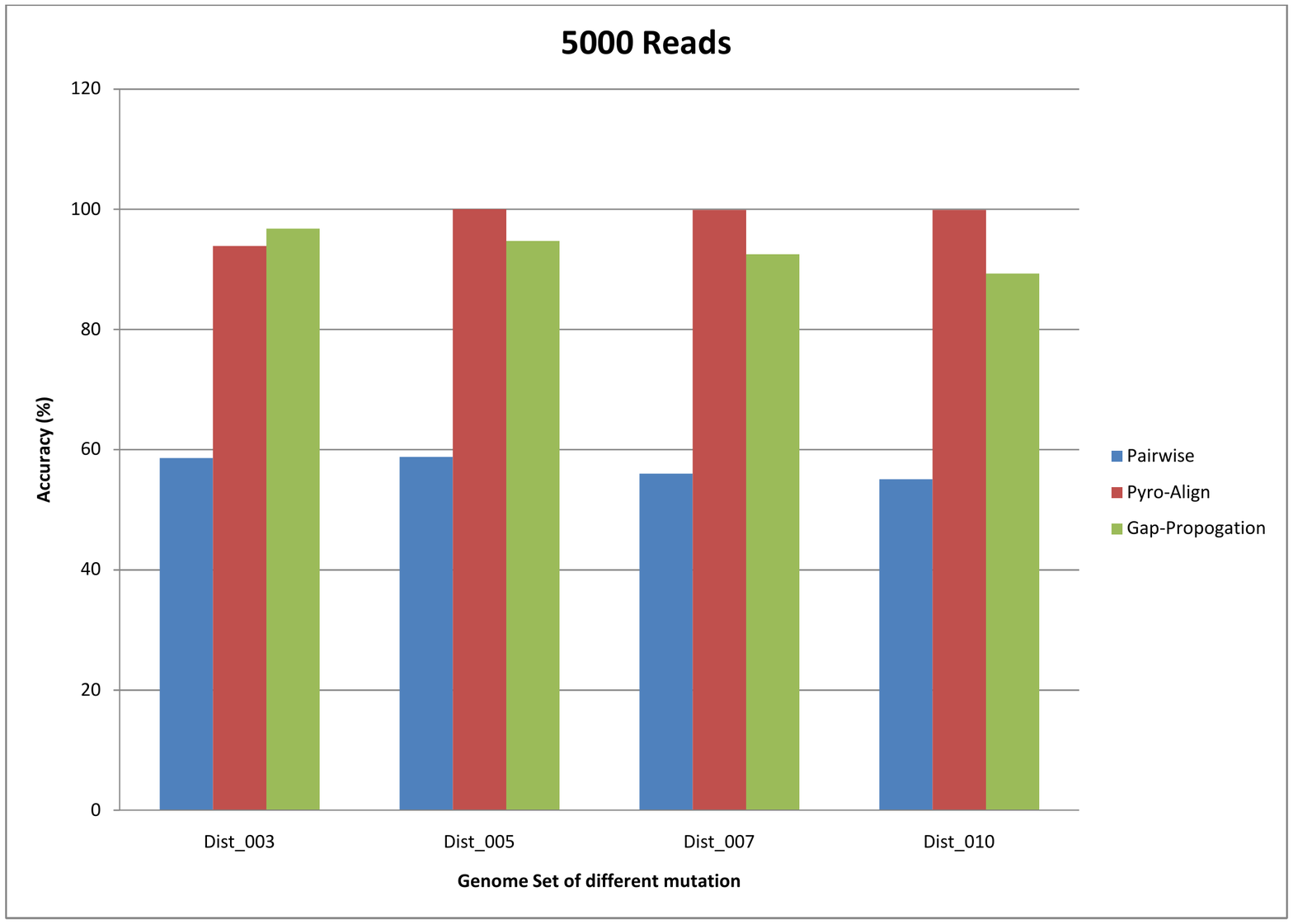}\vspace{-8mm}
\caption{\small \label{fig-5000} The quality of the alignment
using pairwise, pyro-Align and 'propagation' methods for 5000
reads} \vspace{-8mm}
\end{center}
\end{figure}

After the alignment, a majority consensus of the reads is
obtained. A distance based similarity is then calculated of the
consensus obtained from the aligned reads with the original genome
from which the reads were generated.The results of the alignment
obtained and the accuracy of the consensus thus obtained are shown
in Fig.~\ref{fig-2000} and Fig.~\ref{fig-5000} for 2000 and 5000
reads respectively.

We compare the accuracy of the algorithm with two different
methods. First being the simple pair-wise alignment of the reads
with the reference genome. Secondly, we compare it with a
sequential gap propagation method, used in recent pyrosequencing
systems \cite{niko}. Simply put, gap propagation method builds
multiple alignment from pairwise alignments by sequentially
'propagating' the gaps from each pairwise alignment to all the
reads in the system. Propagation of gaps is accomplished for every
position where at least one read has an inserted base. A gap is
inserted in the reference genome and, consequently, in all reads
that overlap the genome at that position. The complexity of the
procedure is of the order of $O(N^2)$.

The accuracy of the consensus obtained using just the pairwise
alignment is less than $55\%$ and that obtained from the
pyro-align is always greater than $96\%$.An even better alignment
quality is achieved for greater number of reads, because more
number of reads provide a better coverage for a genome of given
length. The accuracy of the gap propagation procedure, is
comparable to pyro-align for small mutations, but as the mutations
increase the accuracy of gap propagation based method decreases.

\begin{figure}[htb]
\vspace{-8mm}
\begin{center}
\includegraphics[scale=0.4,angle=0]{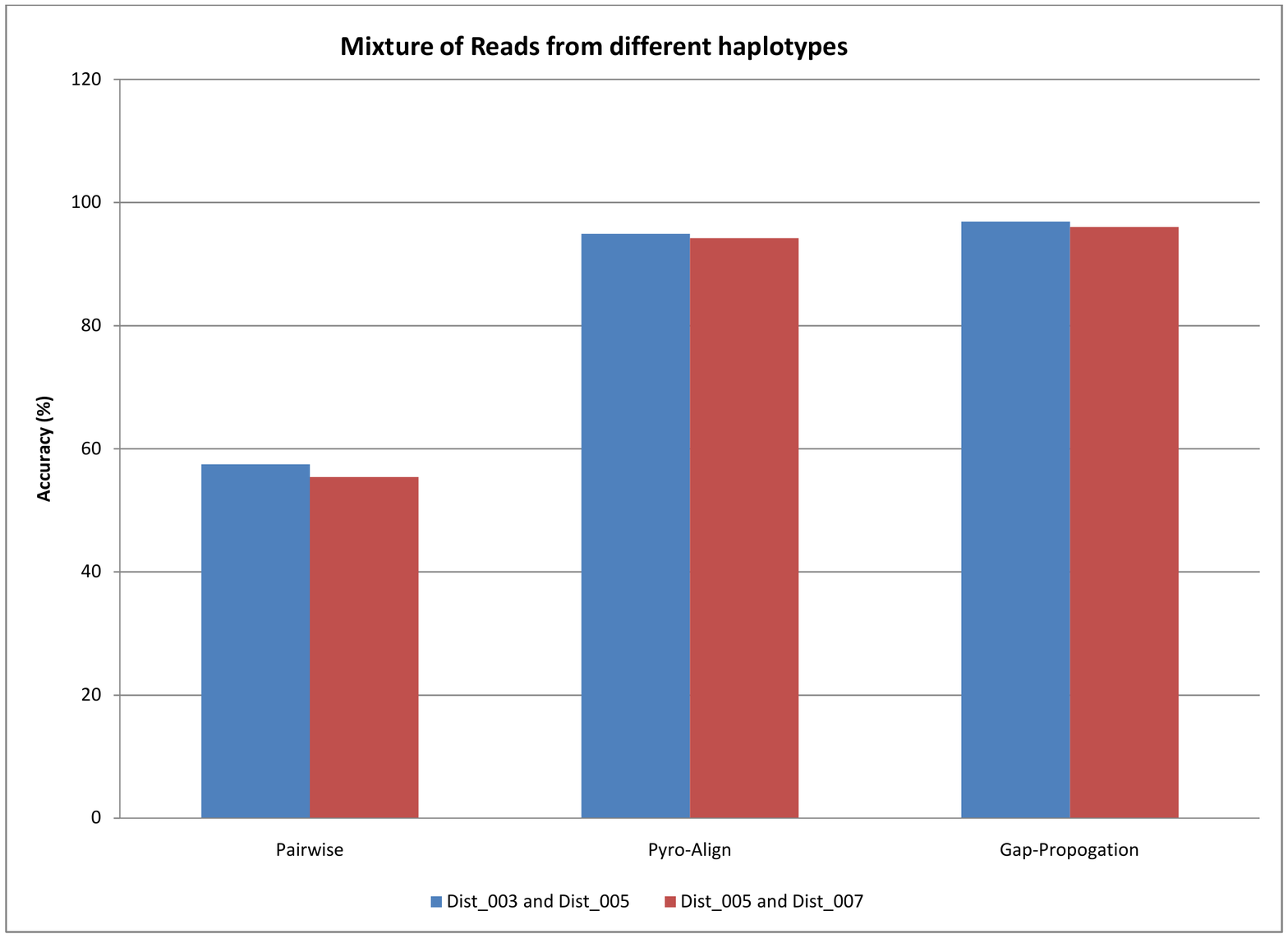}\vspace{-8mm}
\caption{\small \label{fig-mixture} The quality of the alignment
using pairwise, pyro-Align and 'propagation' methods for mixed
genome reads} \vspace{-8mm}
\end{center}
\end{figure}

To illustrate that the alignment system also works with a
'mixture' of reads from different haplotype, we use the mutated
reads from Dist-003, Dist-005 and Dist-007 to generate a new set
of reads. The new set contains equal number of reads from the
mutated sets e.g. 2000 reads from each mutated genome for the
results shown. The reads are then aligned by the pyro-align
algorithm using wildtype as the reference genome. The results of
alignment for this mixture set are shown in Fig.~\ref{fig-mixture}
for Dist-003/Dist-005 and Dist-005/Dist-007 mixtures. It must be
noted here that we don't have a 'ground truth' genome in these
cases and hence no genome is available to compare the consensus
obtained from the alignment.

However, we do know the mutation rates for the genomes from which
the mixture sets were generated. Therefore, if an optimal
alignment of these reads is obtained, the 'mutation' in the
consensus should not be greater than the combined mutations of the
genomes. For example consider the case of Dist-003/Dist-005
mixture. We know the mutation rates for the genomes from which the
reads generated were $3\%$ and $5\%$ with respect to the wildtype.
Therefore, for accurate alignment, the consensus of the alignment
should not vary more than $8\%$, in the worst case, when compared
to the wildtype. Same would be true for the other cases considered
according to the mutation rates of the genomes. As can be seen
that the results of the alignment compared with the wildtype are
well within the expected limits. The accuracy of the pairwise
alignment of the reads with the reference genome(in this case the
wildtype), and that obtained using propagation method is also
shown for comparison.

\section{Complexity Analysis}

In this section we briefly outline the complexity of the proposed
pyro-align algorithm. Recall the pyro-align algorithm consists of
these major steps: semi-global alignment, reordering, pair-wise
alignment, and profile-profile alignment.

We assume that the number of reads is $N$ with the average length
of the read equal to $L_R$. Let's further assume that the length
of the reference genome is equal to $L_g$. Then, the complexity of
the semi-global alignment (overlapping alignment) is equal to
$O(NL_RL_g)$. The clustering of the reads can be done in $O(NL_g)$
and the reordering using hashtables can be achieved in $O(N)$,
making the total for this stage equal to $O(NL_g+N)$. The pairwise
alignment of the reads is shown to be achieved in $O(NL_R^2)$ and
the profile-profile alignment can be achieved in $O(N logN \times
L_g^2)$. This makes the total complexity equal to $O(NL_RL_g +
NL_g + N + NL_R^2 + N logN \times L_g^2)$. This is asymptotically
equal to $O(N log N \times L_g^2 + NL_R^2)$.The advantage of low
complexity of pyro-align was further evident by our
experimentation. We were able to align 2000 reads of average
length $250 bp$ from a genome of length 1970bp in about 12 minutes
compared to 6 hours of computation using more traditional multiple
alignment systems such as Clustalw.

\section{Conclusion}\label{discussion}
\vspace{-5mm} The short reads from the pyrosequencing method are
rendered useless if they are not multiple aligned for magnitude of
important applications, such as haplotype reconstruction and error
elimination. We have presented an efficient hierarchical procedure
to multiple align large number of short reads from the
pyrosequencing procedure.

We demonstrated that simple-pair-wise alignment is not feasible in
the case of pyroreads. We also showed that the proposed method is
much faster than traditional time consuming multiple alignment
methods such as Clustalw or Tcoffee. We also presented the quality
assessment results and compared those with the results obtained by
simple pair-wise alignment procedure and 'propagation' methods.
\vspace{5 mm}

\textbf{Acknowledgements:}The work was done, in part when Fahad
Saeed was visiting Biosystems Science and Engineering Department
(D-BSSE), ETH Zurich, Switzerland and in part, at Department of
Electrical and Computer Engineering, University of Illinois at
Chicago. Saeed was additionally supported by ThinkSwiss
Scholarship by the government of Switzerland.

\end{document}